\documentclass[useAMS,usenatbib]{mn2e}
\usepackage{times}
\usepackage{rotating}
\usepackage{subfigure}

\def \kms{\ifmmode{~{\rm km\,s}^{-1}}\else{~km~s$^{-1}$}\fi}
\def \vhel{\ifmmode{V_{{\rm hel}}}\else{$V_{{\rm hel}}$}\fi}
\def \vsys{\ifmmode{V_{{\rm sys}}}\else{$V_{{\rm sys}}$}\fi}
\def \vobs{\ifmmode{V_{{\rm obs}}}\else{$V_{{\rm obs}}$}\fi}
\def \degree{\ifmmode{^{\circ}}\else{$^{\circ}$}\fi}
\def \lsun{\ifmmode{{\rm\ L}_\odot}\else{${\rm\ L}_\odot $}\fi}
\def \msun{\ifmmode{{\rm\ M}_\odot}\else{${\rm\ M}_\odot$}\fi}
\def \myr{\ifmmode{{\rm\ M}_\odot{\rm\ yr}^{-1}}\else{${\rm\ M}_\odot$ 
yr$^{-1}$}\fi}
\def \teff{\ifmmode{{\rm{T}}_{\rm eff}}\else{${\rm{T}}_{\rm eff}$}\fi}
\def \mdot{\ifmmode{{\rm\dot{M}}}\else{${\rm\dot{M}}$}\fi}

\newcommand{\ha}{H$\alpha$}

\newcommand{\niiab}{[N\,{\sc ii}]\ 6548,\ 6584\,\AA}

\newcommand{\sii}{[S\,{\sc ii}]}
\def \st{\ifmmode{^{\mathrm{st}}}\else{${^{\mathrm{st}}}$}\fi}
\def \nd{\ifmmode{^{\mathrm{nd}}}\else{${^{\mathrm{nd}}}$}\fi}
\def \rd{\ifmmode{^{\mathrm{rd}}}\else{${^{\mathrm{rd}}}$}\fi}
\def \th{\ifmmode{^{\mathrm{th}}}\else{${^{\mathrm{th}}}$}\fi}
\newcommand{\hnii}{{\rm H}$\alpha+$[N~{\sc ii}]}

\title[The expansion proper motions of the giant lobes of PN KjPn~8]
{The expansion proper motions of the extraordinary giant lobes 
of the planetary nebula KjPn~8 revisited}
\author[Boumis \& Meaburn]{P. Boumis$^{1}$\thanks{E-mail:
ptb@astro.noa.gr}, J. Meaburn$^{2}$\\
$^{1}$Institute of Astronomy, Astrophysics, Space Applications 
and Remote Sensing, National
Observatory of Athens, I. Metaxa \& V. Pavlou, P. Penteli, GR-15236
Athens, Greece\\
$^{2}$Jodrell Bank Centre for Astrophysics, University of Manchester,
Oxford Rd., Manchester, UK. M13 9PL.\\
}

\begin{document}

\date{}

\pagerange{\pageref{firstpage}--\pageref{lastpage}} \pubyear{2013}

\maketitle

\label{firstpage}

\begin{abstract}

The primary aim is to establish a firm value for the distance to the
extraordinary planetary nebula KjPn~8. Secondary aims are to
measure the ages of the three giant lobes of this object as well as
estimate the energy in the eruption, that caused the most energetic outflow,
for comparison with that of an intermediate luminosity optical transient 
(ILOT).
For these purposes a mosaic of images in the \hnii\ optical
emission lines has been obtained with the new Aristarchos telescope 
in 2011 for comparison with the
images of the KjPn~8 giant lobes present on the POSSI--R 1954 and POSSII--R
1991 plates. Expansion proper motions of features over this 57 yr
baseline in the outflows are present.
Using these, a
 firm distance to KjPn~8 of 1.8 $\pm$ 0.3 kpc has been derived for now
the angle of the latest outflow to the sky has been established
from HST imagery of the nebular core. Previously, the uncertain 
predictions of a bow--shock
model were used for this purpose.
 The dynamical ages of the three separate outflows
that form the giant lobes of KjPn~8 are also directly measured as 3200,
7200 and $\geq$ 5$\times10^{4}$ yr respectively which confirms their sequential
ejection. Moreover, the kinetic energy of the youngest and most energetic of 
these is measured as $\approx$10$^{47}$ erg which is compatible
with an ILOT origin.

\end{abstract}


\section{Introduction}

Three, short--lived, ejection events over a long period 
of time, and along different ejection axes, must have created the point
symmetric ionized knots (A1 \& A2, B1 \& B2 and C1  marked in Fig. 1 (C2
 may exist but be obscured by foreground dust)
in the giant (14 $\times$ 4 arcmin$^{2}$) lobes 
\citep{lop95} which project from the otherwise
innocuous planetary nebula KjPn~8. The latter forms a bright
4 $\times$ 2 arcsec$^{2}$ elliptically--shaped ring of optical emission
surrounding a star of as yet unknown type \citep{lop00}.
This central compact ring is the radiatively ionized inside surface of a small
central hole in a much 
larger, $\approx$ 20 arcsec diameter, and massive, 0.007 \msun,
disk of neutral molecular gas (\citealt{hug97}; \citealt{for98}).

Some of the characteristics of KjPn~8 and its giant lobes 
suggest their formation, similar to that of the poly--polar planetary
nebula NGC 6302 \citep{sok12}, by an
intermediate--luminosity optical transient (ILOT) after
the eruption identified in NGC 300 \citep{pri09}. 
A precursor to this transient outburst is identified as a dust enshrouded
$\approx$~10\msun\ star and, with a peak luminosity of M$_{V}$ $\approx$
-13, the explosion is identified as an event with lower energy than a 
typical supernova explosion.
The
several months long outburst of an ILOT 
is thought to be powered by mass accretion onto a main
sequence companion to an AGB (or extreme AGB) star.
 In order to form
the point symmetric knots at different orientations in KjPn~8 the rotation
axis of such an interacting 
binary star would have to have changed in time. Certainly,
the point--symmetric knots (A1 \& A2 in Fig. \ref{fig1}) 
which are aligned perpendicularly
to the ellipse of the core of KjPn~8 have the ejection velocities in the 
hundreds of \kms\ \citep{lop97} and could qualify for  this ILOT origin.

Crucial to any interpretation of the nature and origin of KjPn~8 is sound 
knowledge of its distance, D. This has been given as D $=$ 1600 $\pm$\ 230 pc
by Meaburn (1997 -- Paper I)
from measurements of the expansion proper motion (EPM) of 
the A1 \& A2 outflow. This initial, but limited, 
measurement was made for only one sub--knot
in this outflow by comparing its position on a 1954 Palomar Observatory
Sky Survey (POSSI--R) plate to that on a comparable 1991 POSSII--R plate (i.e.
over a 37 yr baseline).

This baseline has now been extended to 57 yr by obtaining a new \hnii\ 
image in 2011 for comparison with the earlier POSS images. This
new \hnii\ image has not only
permitted  more detailed and accurate 
measurents of the EPMs of the A1 \& A2 outflow,
but has also led to the determination of 
their hitherto unknown values for the B2 and C1 features. Newly
excited compact knots are also found in the most recent \hnii\ image
of the A1 \& A2 outflow. Any confusing star images have now been positively
identified by taking a complimentary image in the adjacent red continumm light 
devoid of emission lines.
 
Furthermore, the tilt of the A1 \& A2 outflow to the plane of the sky has
now been 
estimated more certainly  than in Paper I.
When combined with this outflow's measured radial velocity a more rigorous 
outflow tangential velocity, crucial to the determination of D from
the measured EPMs, has now been obtained.
In the present paper, because the measured 
\citep{lop97} expansion velocity of the principal 
lobes is $\approx$ 300 \kms, it is assumed that collisional ionization
by hypersonic shocks must generate the optical emission lines. The high--speed
knots and filaments that constitute these lobes have therefore been 
ejected from the central source. Any complication to the present 
measurements of the EPMs caused
by any radiative ionization by Lyman photons from the central stellar system
must be neglible, for bright rims associated with such ionization
fronts would only propagate into the ejected material 
at the sound speed of $\approx$ 10 \kms. 
A very similar
situation occurs in the measurements of the lobe EPMs of the comparable
multi--polar planetary nebula NGC 6302
\citep{mea08} where the measured outflow velocities are $\leq$ 600 \kms.

\section{Imaging}

\subsection{Observations}

KjPn~8 and its giant lobes were imaged at the f/8
Ritchey--Chretien focus of the 2.3-m Aristarchos
telescope at the Helmos Observatory, Greece on August 18, 2011.
Three pointing positions were used to cover the whole object
with the 5 $\times$ 5 arcmin$^2$ field of view of the CCD detector 
with its 1024 $\times$ 1024 data taking windows each 24 $\mu$m 
square ($\equiv$ 0.28 arcsec square).

\begin{figure*}
\centering
\includegraphics[scale=0.9]{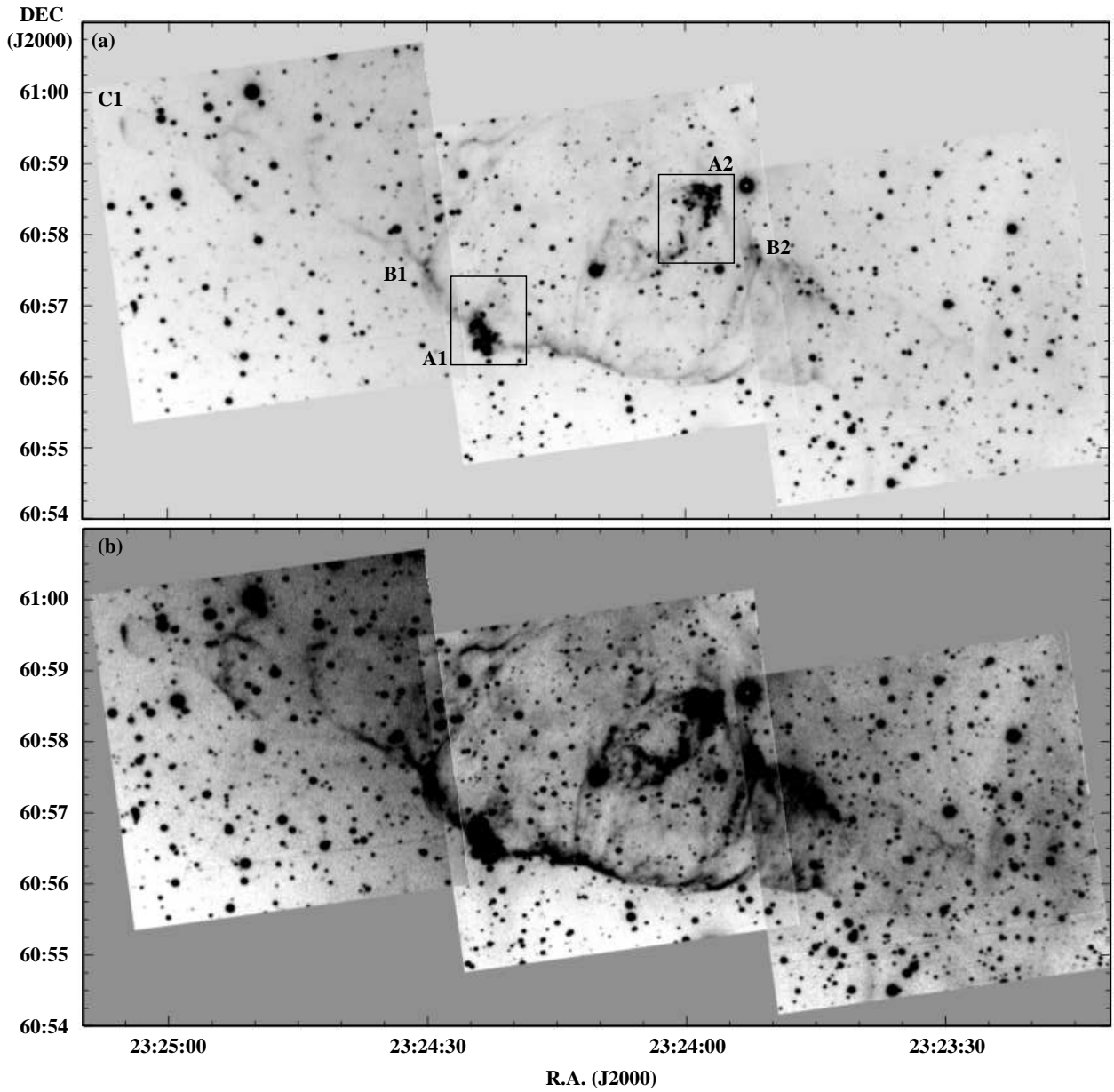}
\caption{a) A light, negative greyscale presentation of mosaic of \hnii\ 
images
of KjPn~8 and its filamentary lobes that were
obtained in 2011 with the Aristarchos telescope is shown. The ionized knots 
A1 \& A2, B1 \& B2 and C1
are indicated. The two rectangular boxes are the areas whose
contours of equal surface brightness are shown in Figs. \ref{fig3} and 
\ref{fig4} respectively.
b) A deeper presentation of the mosaic in Fig. \ref{fig1} a reveals many 
faint features
in the extended lobes.
}
\label{fig1}
\end{figure*}

\begin{figure*}
\centering
\includegraphics[scale=0.8]{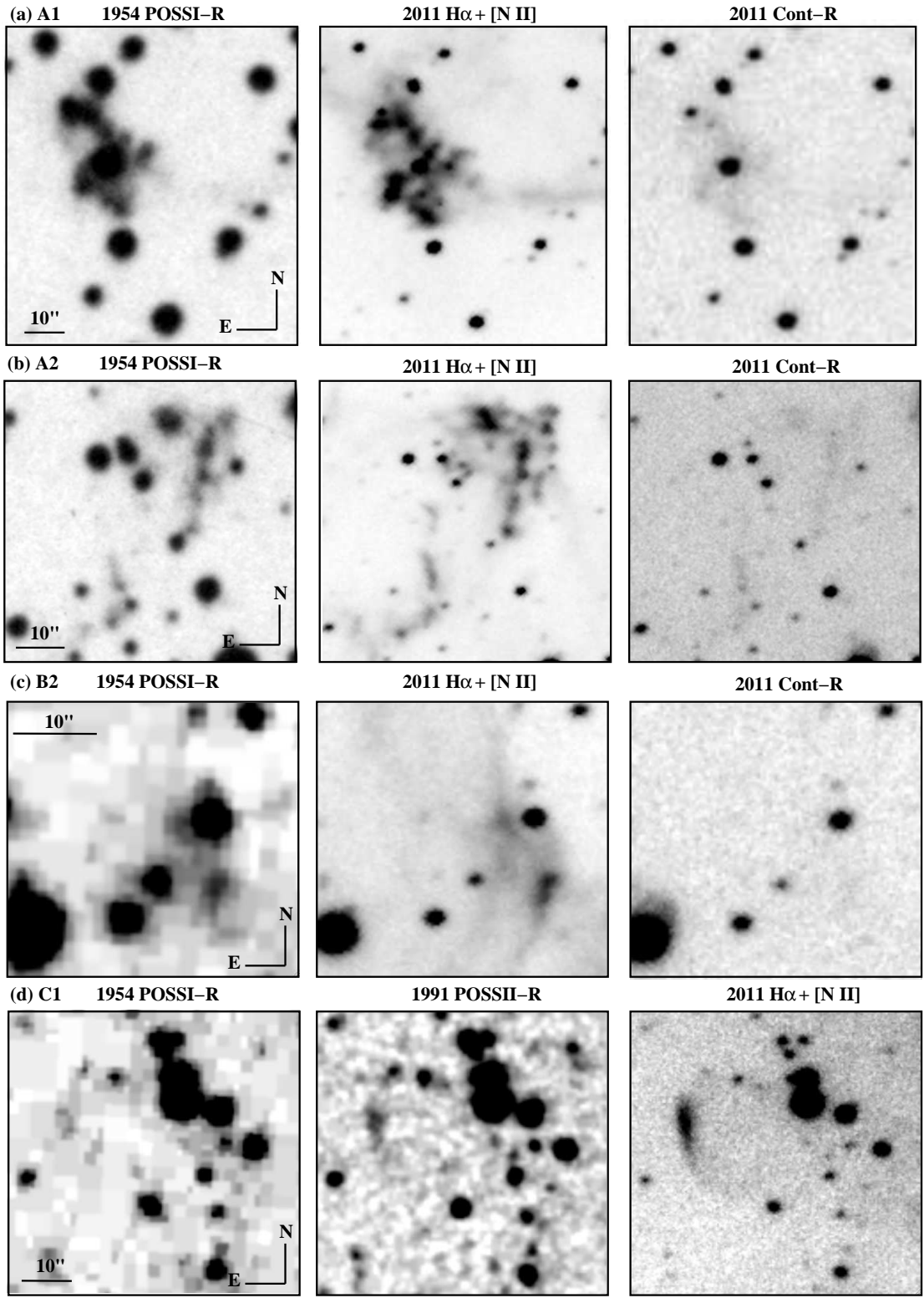}
\caption{a) Images that have been very accurately orientated, scaled and 
aligned with each other of the A1 group of knots (Figs. \ref{fig1}a \& 
\ref{fig3})
are shown. The 
first is the POSSI--R 1954 image to be compared with the latest
2011 images in the lines of \ha\ $+$ \niiab\ and adjacent
continuum light respectively. b) As for Fig. 2a but for
the A2 outflow shown in Figs. \ref{fig1} \& \ref{fig4}. c) As for Fig. 
\ref{fig2}a but for
the B2 feature in Figs. \ref{fig1} \& \ref{fig6}.
 d) Similar images of the C1 feature in Fig. \ref{fig1} are 
compared from the POSSI--R (1954) and  the POSSII--R (1991)  
surveys with the 
most recent 2011 \hnii\ image.}
\label{fig2}
\end{figure*}

Exposures of 1800 s duration were obtained
through the \hnii, 40 \AA\ bandwidth filter, centred 
on 6578 \AA, on each pointing centre. These were complemented
by exposures in the nearby continuum light each of 180 s duration through
a 100 \AA\ bandwidth filter centred on 6680 \AA. {All fields were 
projected on to a common origin on the sky and were subsequently combined 
to create the final mosaics in \hnii. During the observations the seeing 
varied between 0.8 and 1.2 arcsec. The image reduction was carried out using 
the IRAF and STARLINK packages. The astrometric solution for all data frames 
was
calculated using the Space Telescope Science Institute (STScI) Guide
Star Catalogue II (GSC--II; \citealt{las08}). All the
equatorial coordinates quoted in this work, refer to epoch 2000.

The resulting mosaic of these \hnii\ images
is printed lightly in Fig.\ref{fig1}a and deeper in Fig. \ref{fig1}b.  
The point symmetric knots A1 \& A2 and B1 \& B2,
following \citet{lop95}, are identified. The counterpart (C2?) to C1
if it exists
must be  obscured by foreground dust.

%

\section{Results} 
Small areas of the 2011 \hnii\ images covering the A1 \& A2 and B2 knots are 
compared in Figs. \ref{fig2}a--c respectively with their 1954 POSSI--R
images and red 2011 continuum images. POSS--I and POSSII--R images 
of the knot C1 are compared in Fig. \ref{fig2}d with the present 2011 \hnii\ 
image.

The Starlink CCDPACK programme CCDALIGN was used to scale, position the 
stars and orientate each frame identically in each set of three.
The centroids of faint stellar
images are positioned to  within 0.1 arcsec of each other in each set 
after this process i.e. in each of the resultant three images of the same field
any faint star, which has no significant PM of its own, has nearly 
exactly the same
spatial pixel coordinates. The small differences in angular resolution
between the 1954 POSSI-R and the new  2011 \hnii\ (see
the similar angular size of the 
images of the faint star just to the north-west of Knot
A2i, identified in Fig. 4 and displayed in Fig. 2b) do 
not affect this process.
 Only the ionized knots with significant EPMs 
change position between 1954 and 2011. Moreover, the brightnesses 
of some have changed considerably in this intervening period.

The detailed measurement of an EPM involved contouring
the photographic density of an unresolved nebular knot or fine filament
in the case of a CCDALIGN POSSI-R image for comparison
with the contours of relative surface brightness
in the corresponding CCDALIGN 2011 \hnii\ image of the pair.
The displacement of the peak of a fine knot or ridge of a filament
were simply then measured with a ruler on the two contour maps.
This was repeated independently three times and the conservative errors quoted
in Table 1 represent the spread in values from this process.}

\begin{figure*}
\centering
\includegraphics[scale=0.70]{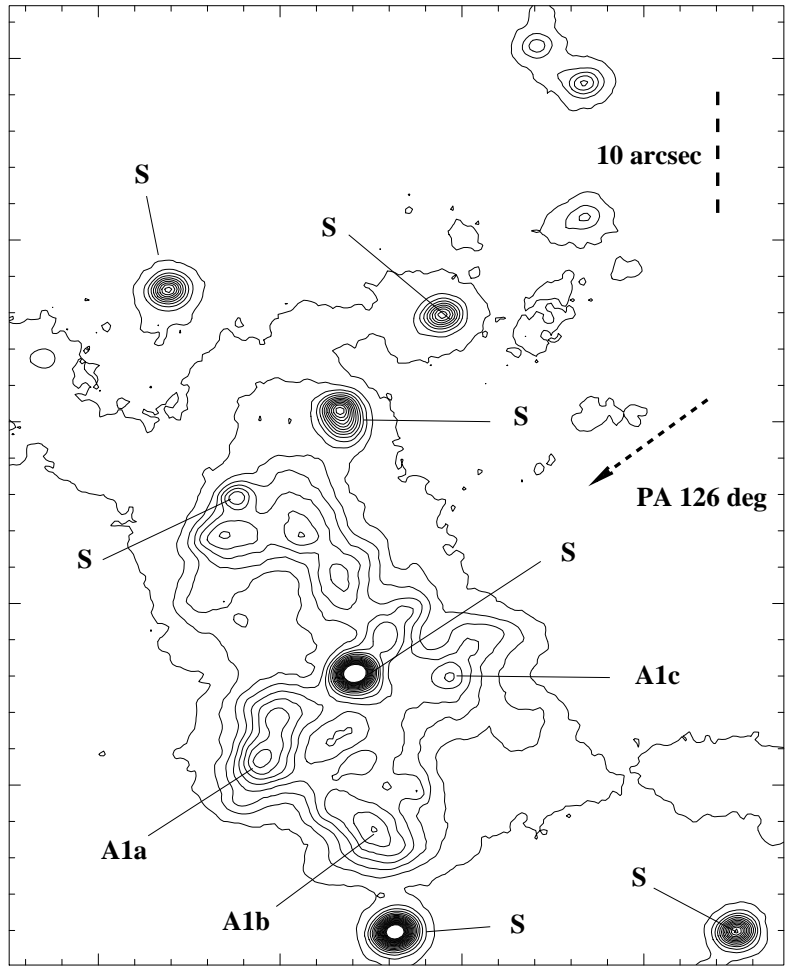}
\caption{The features whose expansive proper motion have been measured
in the A1 group of knots (Fig. \ref{fig1}) are identified in this contour map
of the 2011 \hnii\  image by A1a, A1b etc.
The confusing stellar images are marked by an S. The axis of the central
elliptically-shaped image of KjPn~8 is shown as an arrowed, dashed, line.}
\label{fig3}
\end{figure*}

\begin{figure*}
\centering
\includegraphics[width=\textwidth]{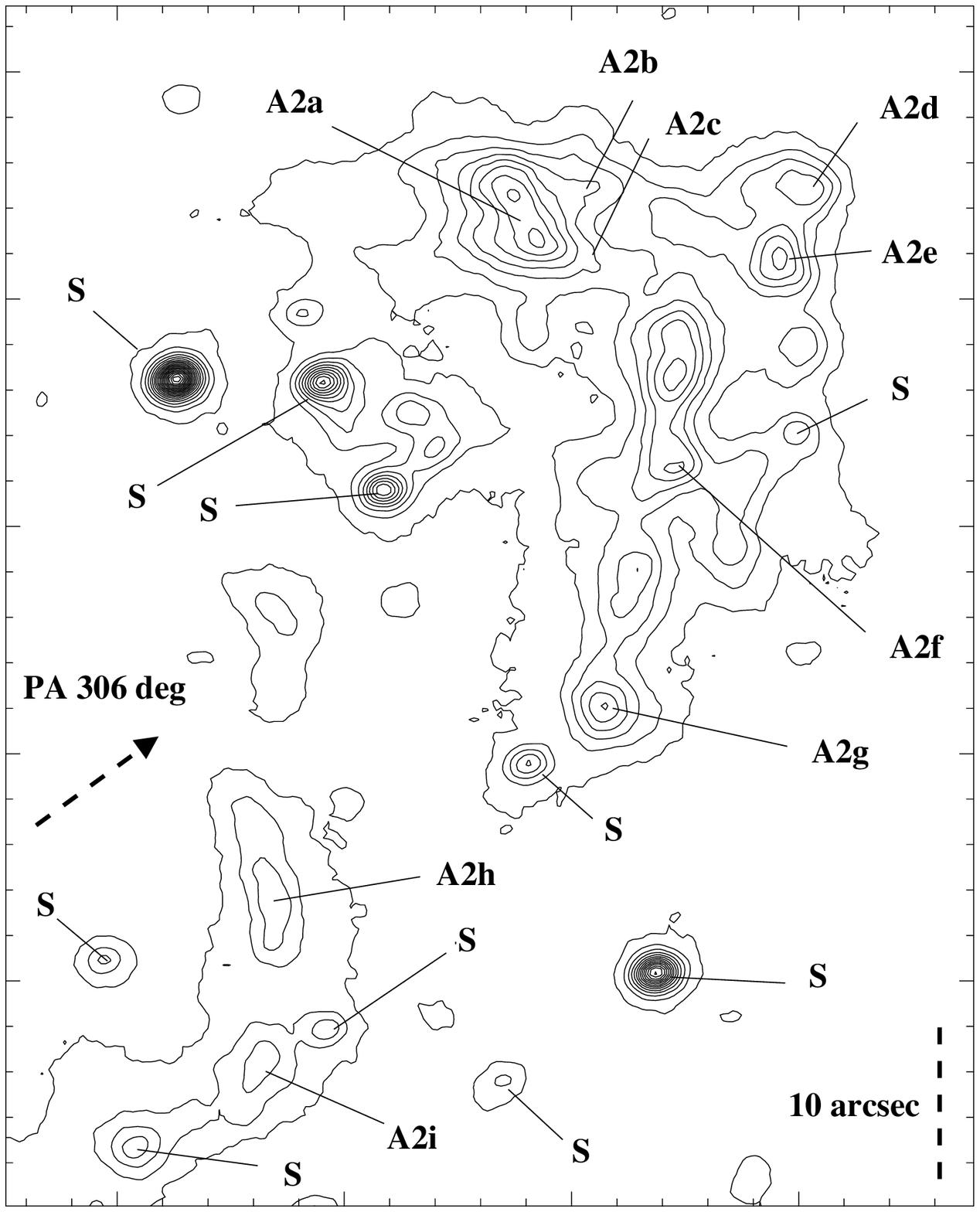}
\caption{As for Fig. \ref{fig3} but for the A2 group of knots identified in
Fig. \ref{fig1}.}
\label{fig4}
\end{figure*}

\begin{figure*}
\centering
\includegraphics[width=\textwidth]{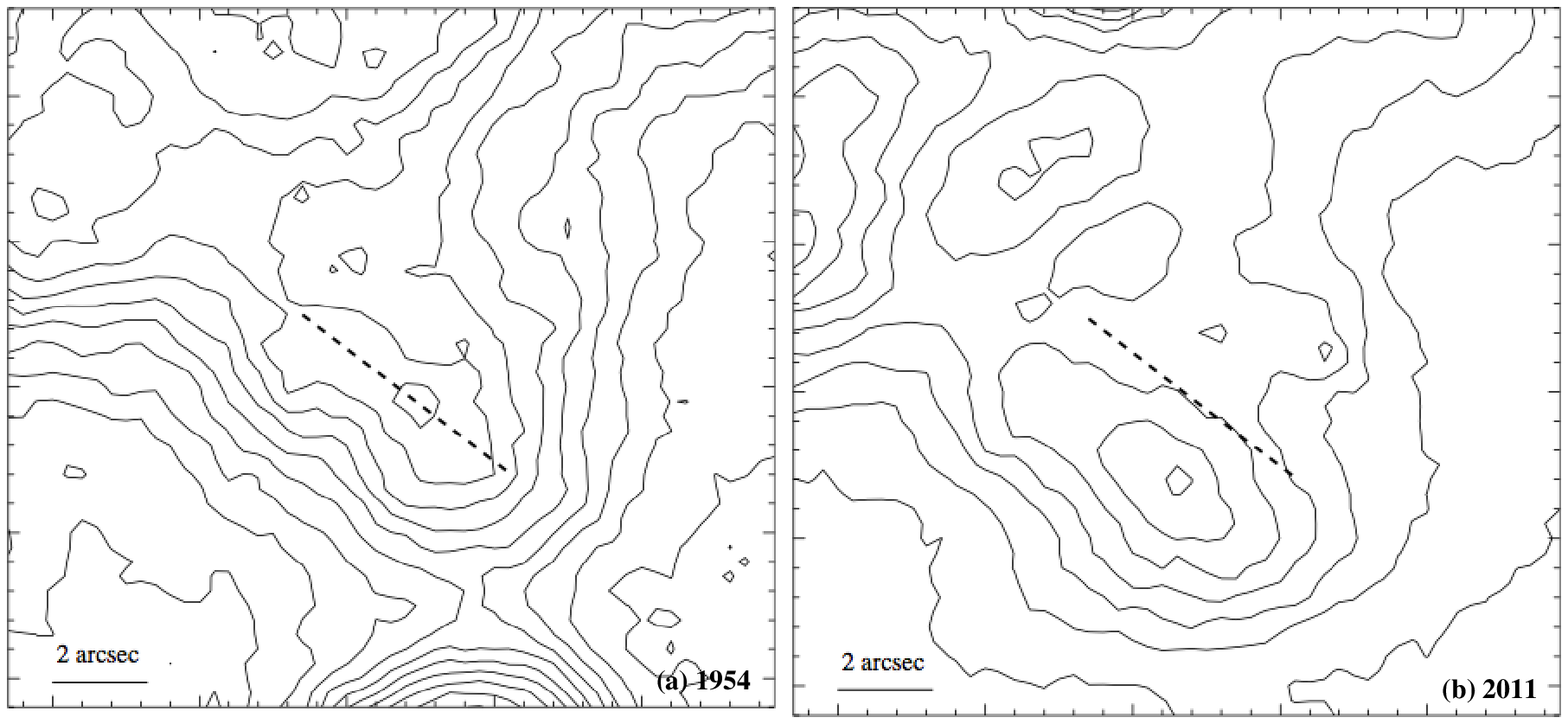}
\caption{a) Contours of the 1954 POSSI--R image from the ccdalign pair 
of Knot A1b identified in Fig. \ref{fig3} and shown in Fig. \ref{fig2}a.
The dashed line indicates the peak
of the ridge of this elongated feature. b) The 2011 \hnii\  
contours of the same knot A1b.
The dashed line is exactly that shown in (a). }
\label{fig5}
\end{figure*}

\begin{figure*}
\centering
\includegraphics[width=\textwidth]{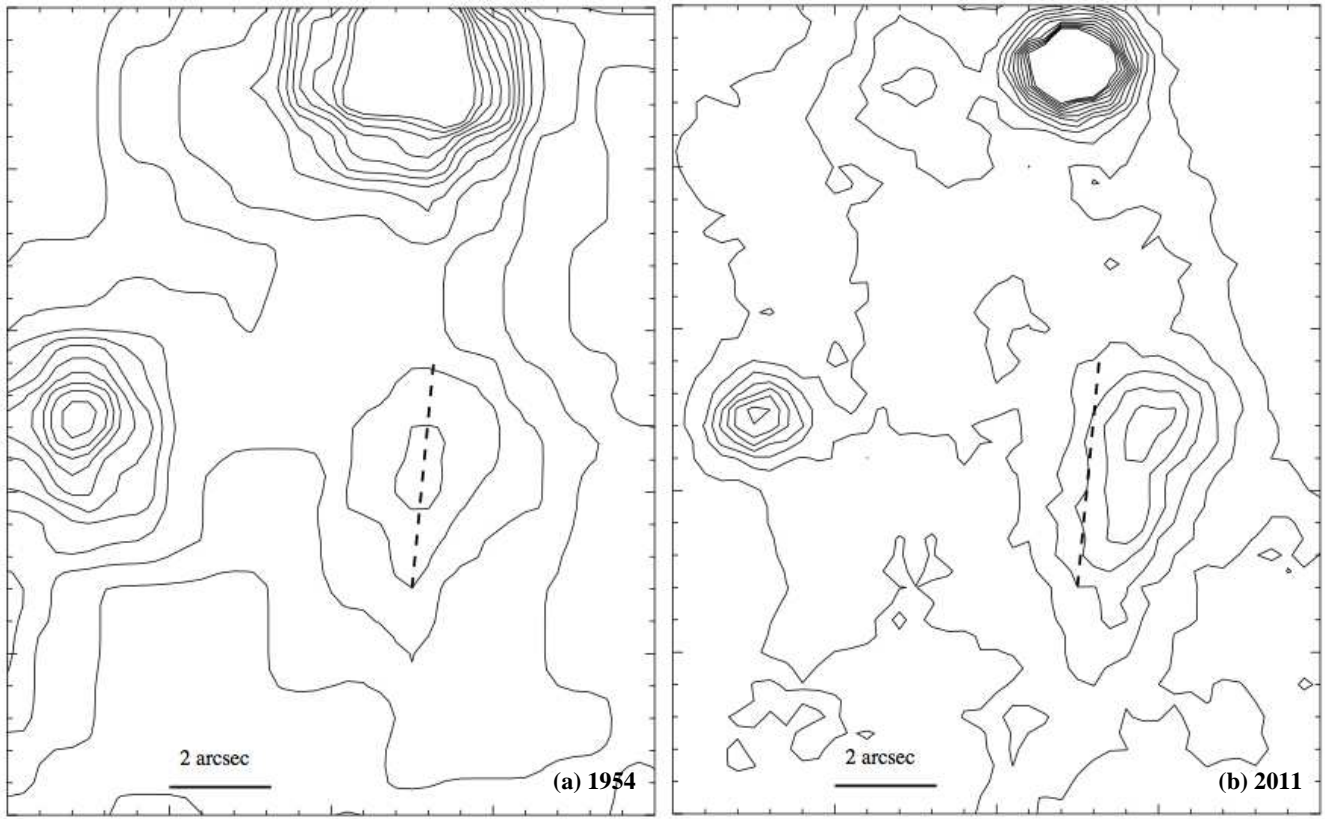}
\caption{a) Contours of the 1954 POSSI--R 
image from the ccdalign pair, of the Knot B2
identified if Fig. \ref{fig1} and shown in Fig. \ref{fig2}c. 
The dashed line indicates the
peak of this elongated feature. b) Contours of the 2011 \hnii\ 
image for the same Knot B2. The dashed line is exactly that shown in (a).}
\label{fig6}
\end{figure*}

The A1 \& A2 feature in Fig. \ref{fig1}a is the  most notable bi--polar 
outflow with
the EPMs very apparent when the 1954 and 2011 images (Figs. \ref{fig2}a and 
b) are
`blinked' using Starlink GAIA software. However, only compact knots 
or fine filaments which appear clearly in both images, separated by 57 yr, 
can be
used to get accurate measurements of these EPMs. The 2011 \hnii\ 
brightness contours of those features that meet this criterion 
(A1a--c and A2a--i) are identified
in Figs. \ref{fig3} \& \ref{fig4} for the A1 and  A2 outlows respectively. 
The axes of the
outflows are also shown and confusing stellar images are marked with an "S". 
Contour
maps of these knots in the  1954 POSSI--R image were compared with those
in the 2011 \hnii\ image to give the EPMs and their position 
angles (PAs) 
to $\pm$5$\degree$ as listed in Columns 2, 3 and 4 of 
Table~\ref{table1}. An example of this
comparison for knot A1b is shown in Figs \ref{fig5}a \&\ b. Only one position 
for
knot B2 was measured in a similar way (Figs. \ref{fig6}a \&\ b) and listed in 
Table~\ref{table1}.
The data for B1 was too poor to permit an accurate measurement.
An upper limit to the EPM for knot C1 is also given. The strange 
knot A2e is only listed with no measurement to draw attention to its 
appearance in 2011. It is not present in the 1954 POSSII--R image 
(see its identification in Fig.4 and presence in Fig. \ref{fig2}b).

The errors in the EPM measurents in Table~\ref{table1} are mainly a consequence
of the changing shapes of the individual knots between 1954 and 2011.
This change is particulary noticeable for knot A2a as can be seen in Fig. 
\ref{fig2}b.

The validity of obtaining the EPMs of expanding lobes by comparison
of nebular images on a  POSSI-R 103a-E photographic plate with those
obtained in 2011 using a CCD detector, as well as a different filter
(with 40 \AA\ bandwidth)
and telescope, should be questioned. The 
\ha\ line at 6563 \AA\ and the \niiab\ nebular emission lines dominate
all others in both observations with intensities relative to each other
in the A1 lobe of 855, 1598 and 528 respectively \citep{vas98}. 
However, the broader
wavelength range ($\approx$ 800 \AA) of the POSSI-R observation
also transmits the \sii\ 6716, 6731 \AA\ nebular
lines whose relative intensities are 548 and 410 respectively. This 
causes neglible problems in this comparison for the sensitivity
of the Kodak 103a-E falls off rapidly towards the red: it is 
around 40 times less sensitive at 6730 \AA\ than at 6563 \AA\
\citep{west50}. In any case, any residual detection of the \sii\ lines,
because   they originate in ions of 
similar ionization potential to those that emit the \niiab\ lines, would 
make no significant difference to a determination of the position 
of a feature in a lobe of  KJPn~8.
The ratio of continuum starlight to the nebular emission line light
is of course much higher in the broad band POSSI-R image than in the
narrower bandwidth 2011 \hnii\ image. This leads to the detection of more
faint stars and the spread in the photographic emulsion of the images of
brighter stars in the POSSI-R data. However, the angular resolution is 
unaffected and very similar in both observations (see the faint stellar
images 
in Fig. 2a-d). With these differences in mind only the changes in positions
of the peak brightnesses of compact knots and fine filaments 
are measured in the present work.

Proof of the validity of this comparison and the simple method of analysis 
is found in \cite{mea08} and \cite{szy11} both dealing
with EPMs of similar high--speed lobes of NGC 6302. In  \cite{mea08},
nebular positions from a recent CCD narrow--band image are compared
with those from a historic broadband image obtained with 103aE photographic 
emulsion and different telescope. The EPMs in \cite{mea08} 
match  closely  their values in the  
complementary measurements
made by  \cite{szy11} employing only Hubble Space Telescope images.

Furthermore, the EPM rate of one bright knot in the A1 lobe in Paper 1 , where
its change in position between the POSSI and POSSII surveys was compared,
is reproduced, within the quoted uncertainties, 
in the present work. An identical
method of analysis to that described here was employed.

The dynamical age of a feature given in Table 1 is its angular
distance from the nebular core, measured using the image in Fig. 1, 
divided by its EPM rate.

\section{Dynamical ages}
The dynamical age of a knot assumes that its measured EPM, over
a limited time interval, has been maintained
during its whole passage from its point of origin to its present position. 
Obviously decelerations are likely to have occurred if it encounters
a significant  ambient medium since leaving 
the ejecting source. However, its value can 
give an indication of the sequence of events and it remains
possible (Sect. 6) that the A1 \& A2 as well as the B1 \& B2 outflows
have travelled through tenuous material in pre--formed large cavities.

Here, the ejecting source is assumed to be the central star at
RA(2000) 23h24m10.408s DEC(2000) 60\degree 57\arcmin 30.61\arcsec  discovered
in the Hubble Space Telescope image. The angular separations listed
in Column 5 of Table 1 are the apparent distances from this star. 
The dynamical ages
of the knots listed in Column 6 of Table 1 are then simply the division 
of these separations by the respective EPM rates in Column 4.

It is clear from these dynamical ages that the A1 \& A2 outflows
are the bi--polar consequence of the same event i.e. they have been ejected
in opposite directions with the same speeds but along the same axis.
Only the Knots A2h and A2i fail to fit this pattern. A2h appears to be moving
at a substantial angle to the general outflow and A2i has a significantly
smaller dynamical age. Excluding these two knots the mean dynamical age
of the A1 \& A2 outflows is 3200 yr.

With a dynamical age of 7218 yr knot B2 is a significantly older
feature than the A1 \& A2 outflows and with an age of 
$\geq$ 5 $\times$10$^{4}$ yr
C1 must be the earliest event in the creation of the lobes surrounding
KjPn~8.
\section{Distance}
There are several serious  
weaknesses in the determination of the distance, D, to KjPn~8 
in Paper I. Firstly, the EPM of only one knot in the A1 \& A2
outflow was derived  from comparison of  POSSI--R (1954) and 
POSSII--R (1991) images. It is sounder to include more knots
in this measurement, as in the present paper,
to avoid the possibility that a single knot
could be unrepresentitive.

Furthermore, the crucial value  of the angle 
to the plane of the sky,  $\theta$,
of the outflow was determined using the predictions of 
the theoretical model of bow--shocks \citep{har87}. 
Firstly, this assumes
that the emission is from bow shocks around multiple ejecta, 
which is most likely though not certain. Within this
model it is assumed that the full width of the line profiles equals
the difference in outflow velocity and the medium it is colliding with.
The error on D
quoted in Paper I includes only that from the EPM measurement and does
not include any uncertainty due to these obvious weaknesses in its derivation.

Here ,these  basic weaknesses have now been addressed: the EPMs of 
many knots within the A1 \&\ A2 outflow  and a direct value of 
$\theta$, from a subsequent HST image of the nebular core, have both
been determined.

From the values in Table~\ref{table1} the weighted mean of the A1 \& A2 
outflow EPM (for 
knots A1a, b \&\ c and A2a ,b, c, f \& g) is 1.93 $\pm$ 0.1 arcsec which in 57
yr gives an EPM rate of 33.9 $\pm$ 0.18 mas yr$^{-1}$.
It seems most probable that the A1 \& A2 outflow is along the axis
of the central toroidal disk of KjPn~8 which appears as 
a 2.4\arcsec $\times$ 1.3\arcsec\ elliptical ring. As this is assumed
to be a circular toroid viewed at an angle 
then its axis must be tilted by $\theta$ = 32.6\degree .   
As the magnitude of the difference in radial velocity $\delta$\vhel = 180 \kms\
of both the A1 \& A2 outflows, with respect to the systemic radial velocity,
then the outflow velocity V $=$ 334 \kms\ and tangential velocity 
V$_{t}$~$=$~281~\kms.

The relationship of D, EPM rate and outflow velocity V along a line at an angle
$\theta$ to the plane of the sky is:

\begin{equation}
{\rm D} ({\rm kpc}) \times {\rm EPM (mas~yr}^{-1}) = 0.2168 \times 
{\rm V} ({\rm km~s}^{-1}) \times \cos(\theta)
\end{equation}

Insertion of the measured values in Eq. 1 therefore gives D $=$ 1.8 $\pm$ 0.3
 kpc. The uncertainty in D is larger than that given by the 
standard deviation of the EPMs 
in Table~\ref{table1} alone for it must also 
include the range of \vhel\ values observed within the 
outflow and the uncertainty in the measurement of $\theta$.

Incidentally, the large projected extents of the A1 \& A2, B1 \& B2 and C1
\& C2 (?) lobes of 2.1 pc, 2.4 pc and 7.4 pc respectively
are confirmed by this new distance measurement.

\section{Discussion}

The dynamical ages in Sect. 4 are the first direct evidence of 
the sequence of events that must have
formed the giant lobes of KjPn~8. Firstly, the lobe culminating in C1 \& C2(?)
was formed by a collimated, but shortlived,  bi--polar ejection (episodic jet) 
$\approx$~5$\times$10$^{4}$ yr ago.
Larger cavities around the path of this jet were then created most likely
by the mechanism described by \citet{ste98}. 
Both the B1 \& B2 and A1 \& A2 bipolar
outflows have their brightest optical emissions where they collide with
the edges of these large cavities 7200 and 3200 yr ago respectively. Within
this interpretation these latter dynamical ages are likely to be the real ages
because the ejected material will have travelled along different axes but
through the low density ambient medium
within the evacuated cavity interior. 

The distance measurement to KjPn~8 of D$=$1.8 $\pm$ 0.3 kpc derived in Sect. 5
is on a far firmer footing than the previous one in Paper I. However,
as its value in Paper I is close to this latest determination, the 
bow-shock interpretation of the full width of the line profiles 
(taken to be equal the outflow velocity-- \citealt{har87}) that was  
employed in Paper I, but not here, is vindicated.
It is also interesting that the 15\arcsec\ radius disk
of CO emission surrounding the small optical core of KjPn~8 has an expansion
velocity of 7 \kms\ \citep{for98} to give a dynamical age 
of 2 $\times$ 10$^{4}$ yr
when combined with this latest measurement of D. 
This implies  that the creation of this central neutral disk not only 
certainly preceded  the A1 \& A2 bipolar 
ejection but also, marginally, that of B1 \&  B2.

A sound order--of--magnitude estimation of the kinetic energy (KE)
of the outflows from KjPn~8 can now be made with D$=$1.8 kpc after
the volume of the ionized gas is estimated for the most energetic A1 \& A2 
bipolar outflow.
With the  local electron density 
N$_{e}$ $=$ 100 cm$^{-3}$ from the \sii\ ratios measured by \citet{vas98} 
the outflowing mass of A1 \& A2 ionized gas is then 0.08 \msun\ and 
with V$=$334 \kms\ 
(Sect. 5) the KE $\approx$ 10$^{47}$ erg is derived. Both this KE and V are 
then 
comparable to those for the poly--polar planetary nebula, NGC 6302, 
which led \citet{sok12} to suggest an ILOT origin.
The older lobes of KjPn~8 defined by the C1 \& C2(?) and B1 \& B2 
point symmetric
features (see Table 1) may also have been generated 
by the same binary system  with a rotating
axis  which culminated in the final ILOT outburst of which the A1 \& A2
outflow is the consequence.
 The three
axes of ejection of the KjPn~8 lobes, with different PAs, suggest alone
their origin in a close binary system.
 Another criterion for 
this classification by \citet{sok12}
in NGC 6302 is the Hubble-type nature of its outflowing lobes i.e.
the outflow velocities along a lobe are proportional to the distance
from the origin (see \citealt{mea05, mea08}; \citealt{szy11} for NGC 6302). 
There is only marginal evidence in Table 1 for this to be
the case for KjPn~8. Perhaps,  slower
'bullets'  within the A1 \& A2 outflow 
are travelling within a cavity and not excited until they
hit its boundary where they become excited knots. 
The recent appearance of Knot A2e, seen first 
in the 1991 image (Paper I) and here in the 
2011 image
but not in the 1954 one (Fig. \ref{fig2}b), suggests that this could be 
the case.

The detailed nature of the ionized knots in the A1 \& A2
outflow, however, remains uncertain. Is the knotty optical line 
emission from concave 
bow--shocks, as seen from the central star of KjPn~8, around `bullets',
sprayed out in a wide bipolar cone,
as they encounter the ambient medium or, alternatively, from  convex 
bow--shocks as a bipolar cone of less collimated, ejected, 
material encounters
dense clumps in the surrounding shell wall? The first possibility
seems most likely though the string of knots
coincident with the wall of the larger cavity of A1 in Fig. \ref{fig3}, 
of which
Knot A1c is one of these, could suggest 
the latter possibilty is correct. Images of the knots
 in the optical emission
lines, at higher angular resolution ($\leq$ 0.1\arcsec), are now needed
to determine the concave or convex nature of the bow--shocks.

The ILOT origin of the youngest A1 \& A2 lobes of KjPn~8 
would require an AGB star in a close binary system
at the core of KjPn~8. A candidate for such a star (or its immediate 
AGB aftermath) must be the
MERLIN 6 cm point ($\leq$ 0.1\arcsec\ diameter) thermal radio source
which, though somewhat offset from the centre of the ionized ring of KjPn~8,
is embedded in a larger knot of \ha\ emission within the nebular
core. This object at 
RA(2000) 23h24m10.399s DEC(2000) 60\degree57\arcmin30.14\arcsec\ 
is separated by 0.34\arcsec\ from the HST
central star which corresponds, with D = 1.8 kpc, to an apparent separation of 
800 AU whereas \citet{sok12} give a requirement of $\leq$ 5 AU
for the  mass transfer to occur in the central system of NGC 6302 
within an ILOT configuration. A more detailed investigation 
of the stellar systems 
in the core of KjPn~8, particulary any assocated directly with the MERLIN
radio source, are required before firm conclusions can be drawn. The MERLIN
source could be the required close binary but within a 
more complex stellar system.

\section{Conclusions}

\noindent 1) The EPM rates of the A1 \& A2, B2 and C1 giant lobes of
KjPn~8 are measured as 33.9, 19.3 and $\leq$ 8.8 mas y$^{-1}$ respectively.
The dynamical ages of these features are then 3200, 7,218 and $\geq$ 5 $\times$
10$^{4}$ y respectively.

\noindent 2) The outflow velocity of the A1 \& A2 bipolar lobe is 334 \kms\ 
as derived from its measured changes in radial velocity
and its angle of tilt with respect to the plane of the sky derived from 
HST imagery.

\noindent 3) A very sound distance, D, to KjPn~8
is derived in the present paper
as 1.8 $\pm$ 0.3 kpc. This has been derived by combining the EPM rate
of the A1 \& A2 outflow, now for several unresolved knots, with
the measurement of the radial velocity difference and the newly
derived outflow angle, $\theta$, to the plane of the sky.

\noindent 4) This new value of D is very similar to the less
certain one in Paper I which was derived from the EPM rate 
of only one knot in the A1 \& A2 outflow combined with a more uncertain value
of $\theta$ which depended on the theoretical predictions of a bow shock 
model. This similarity could suggest that the A1 \& A2 line emission is 
originating in bow shocks and the theoretical predictions of the bow shock 
model are sound. Imagery of the A1 \& A2 lobe at 10 times higher angular
resolution could confirm the existence of bow shocks around small high--speed
bullets.

\noindent 5) A candidate for the dust enshrouded massive AGB 
star (or its immediate aftermath)
that is required, if the youngest, and most energetic,
A1 \& A2 lobes of KjPn~8, originated in an ILOT event, could be 
the compact radio source in the central nebula. Any evidence that this
is a close binary system should be searched for.
The older C1 \& C2 (?) and B1 \& B2 lobes could then have been
generated by less energetic ejections, from the same binary system, that
preceded the final ILOT event.

Further consideration of ILOTs generally are outside
the scope of the present paper.

\section*{Acknowledgements}
JM is grateful to the hospitality of the National Observatory of
Athens in June 2012 when this paper was initiated.  The authors would 
like to thank C. Goudis for all the effort he put the last decade in order 
the "Aristarchos" telescope to be operational. 
Based on observations made with the Aristarchos telescope 
operated on Helmos Observatory by the Institute of Astronomy, 
Astrophysics, Space Applications and Remote Sensing of the National 
Observatory of Athens.

\clearpage

\begin{table}
\centering
\caption[]{In Column 1 the ionized knots identified in Figs. \ref{fig1}a, 
\ref{fig3} and \ref{fig4} are 
listed. Their expansion proper motions (EPMs) and position angles (PAs) 
of their motions since 1954 are given in Columns 2 and 3 respectively. The 
rates of these expansive motions over the 57 yr between observations
and their distances from the central star
are given in Columns 4 and 5 respectively. These rates and distances are
converted into dynamical ages in Column 6. The percentage errors listed in
Column 2 pass through to the values in Columns 4 and 6.}
\label{table1}
\begin{tabular}{|c|c|c|c|c|c|}
\hline
\multicolumn{6}{|c|}{Knot Parameters}\\
\hline
Knot & EPM & PA & EPM rate & Separation & 
dynamical age \\
\hline
&arcsec&degrees&mas yr$^{-1}$&arcsec&yr\\
\hline
A1a&2.05$\pm$0.3&146&35.9&118&3284\\
A1b&1.77$\pm$0.2&136&31.6&115&3630\\
A1c&2.85$\pm$0.5&147&50.0&106&2116\\
A2a&2.40$\pm$0.5&317&42.1&110&2606\\
A2b&1.77$\pm$0.3&306&30.5&116&3810\\
A2c&2.13$\pm$0.3&306&37.4&114&3035\\
A2d&2.30$\pm$0.3&315&40.4&126&3121\\
A2e&-&-&-&124&-\\
A2f&1.80$\pm$0.3&300&31.6&111&3525\\
A2g&1.54$\pm$0.3&293&27.2&102&3739\\
A2h&2.96$\pm$0.5&10 &51.9&78&-\\
A2i&2.70$\pm$0.3&336&47.4&75&1577\\
B2&1.10$\pm$0.3&290&19.3&139&7218\\
C1&$\leq$0.5&-&-&423&$\geq$5$\times$10$^{4}$\\
\hline
\end{tabular}
\end{table}

\newpage

%

\end{document}